\font\twelvebf=cmbx12
\font\ninerm=cmr9
\nopagenumbers
\overfullrule=0pt
\magnification =\magstep 1
\line{\hfil CCNY-HEP 98/3}
\line{\hfil RU-98-6-B}
\line{\hfil SNUTP 98-034}
\line{\hfil April 1998}
\vskip .3in
\centerline{\twelvebf On the vacuum wavefunction and string tension of }
\centerline{\twelvebf  Yang-Mills theories in (2+1) dimensions}
\vskip .3in
\baselineskip=14pt
\centerline{\ninerm DIMITRA KARABALI}
\vskip .05in
\centerline{Physics Department}
\centerline{Rockefeller University}
\centerline{New York, New York 10021}
\centerline{karabali@theory.rockefeller.edu}
\vskip .3in
\centerline{\ninerm CHANJU KIM}
\vskip .05in
\centerline{Center for Theoretical Physics}
\centerline{Seoul National University}
\centerline{151-742 Seoul, Korea}
\centerline{cjkim@ctp.snu.ac.kr}
\vskip .3in
\centerline{\ninerm V.P. NAIR}
\vskip .05in
\centerline{Physics Department}
\centerline{City College of the City University of New York}
\centerline{New York, New York 10031}
\centerline{vpn@ajanta.sci.ccny.cuny.edu}
\vskip .3in
\baselineskip=14pt
\centerline{\bf Abstract}
\vskip .1in
We present an analytical continuum calculation, starting from first principles,
of the vacuum wavefunction and string tension
for pure Yang-Mills theories in $(2+1)$ dimensions, extending our
previous analysis using
gauge-invariant matrix variables. The vacuum wavefunction is consistent with
what is expected
at both high and low momentum regimes. 
The value of the string tension is in
very good agreement with recent lattice Monte Carlo evaluations. 
\vfill\eject
\footline={\hss\tenrm\folio\hss}
\baselineskip=18pt
\pageno=2

\def\dag {\dagger}
\def\del {\partial}
\def\bdel{\bar{\partial}}
\def \bd {\bar{\partial}}

\def\d {\delta}
\def\bz {\bar{z}}
\def\half {{\textstyle {1 \over 2}}}
\def\vf {\varphi}
\def\ra {\rangle}
\def\la {\langle}
\def\Tr {{\rm Tr}}
\def\bA {{\bar A}}
\def\bD {{\bar D}}
\def \bE {{\bar E}}
\def \bcalG {{\bar {\cal G}}}

\def \C {{\cal C}}
\def \S {{\cal S}}
\def \vx {\vec{x}}
\def \vy {\vec{y}}
\def \vv {\vec{v}}
\def \vu {\vec{u}}
\def \vk {\vec{k}}
\def \vp {{\vec p}}
\def \vq {{\vec q}}

In recent papers we have done a Hamiltonian analysis of non-Abelian gauge theories
in two spatial dimensions [1,2]. The analysis was facilitated by a special matrix
parametrization for the gauge potentials and the use of some results from
conformal field theory. We obtained results regarding the mass gap
and wavefunctions as well as a reduction of the Hamiltonian to
gauge-invariant degrees of freedom. In this paper, we shall extend our analysis
with a more exact calculation of the vacuum wavefunction and the string
tension. Our results are in very good agreement with recent Monte Carlo
simulations of $(2+1)$-dimensional gauge theories.
It should be emphasized that our work is an analytical calculation
directly in the continuum and based on first
principles. 

We shall begin by a brief recapitulation of the main results.
In our previous papers, we have used the $A$-diagonal representation.
In this paper, we give a reduction of the 
Hamiltonian to the gauge-invariant degrees of freedom in a representation 
independent way, i.e., valid for the $E$-representation as well as 
the $A$-representation, before specializing to the $A$-representation and
recovering the previous result. As far as the kinetic energy operator
is concerned, the vacuum wavefunction is trivially obtained. The effect of the
potential energy is included in a systematic perturbation expansion.
The expansion parameter is $k/m$, where $m={{e^2 c_A} \over 2\pi}$ is the 
mass parameter which emerges from our analysis and $k$ is the characteristic
momentum.
From the vacuum wavefunction, with first order corrections due to
the potential energy, we calculate the expectation value of the Wilson
loop operator. This obeys the area law and gives the value of the string
tension. By summing up sequences of terms in the $1/m$-expansion, the vacuum
wavefunction is reexpressed in terms of a series in $J$, where $J$ is a current
to be introduced below. The terms in this series interpolate smoothly between 
low and high 
momentum (standard perturbative) regimes. Similar analysis for the low energy
excitation spectrum of the Hamiltonian is outlined.

We consider the Hamiltonian version of an $SU(N)$-gauge theory in the
$A_0=0$ gauge. The gauge potentials are $A_i= -it^aA^a_i,~i=1,2$, where
$t^a$ are hermitian $(N\times N)$-matrices which form a basis of the 
Lie algebra of $SU(N)$ with $[t^a,t^b]=if^{abc}t^c,~\Tr (t^at^b)= 
\half \delta^{ab}$. The Hamiltonian can be written as
$$
{\cal H}= T+V,~~~~~~~~~~~~~~~T= {e^2\over 2}\int E^a_iE^a_i ,~~~~~~~~~~~~~~V=
{1\over 2e^2} \int B^aB^a \eqno(1)
$$
where $e$ is the coupling constant, $E^a_i$ is the electric field and 
$B^a = \half \epsilon_{jk}(\partial_jA_k -\partial_kA_j +[A_j,A_k])^a$ is 
the magnetic field. We shall use complex coordinates $z= x_1-ix_2, ~
\bz =x_1+ix_2$ with the corresponding components $A=A_z =
\half (A_1+iA_2), ~ \bA =\half (A_1-iA_2), ~ E=\half (E_1+iE_2), 
~ \bE = \half (E_1-iE_2)$. In (2+1) dimensions, $e^2$ has the dimension of mass.

The parametrization of the gauge potentials we have used in our analysis
is
$$
A= -\partial M~M^{-1},~~~~~~~~~~~~~~~~\bA = M^{\dagger -1}\bdel M^\dagger \eqno(2)
$$
where $M$ is a complex $SL(N,{\bf C})$-matrix.
In terms of this parametrization, the volume element on the space $\C$ of
gauge-invariant configurations can be explicitly calculated
as
$$ 
d\mu (\C )~= d\mu (H)  ~e^{2c_A \S (H)} \eqno(3)
$$
where $H= M^\dag M$ and $d\mu (H)  = \prod _{x}  \det r [d\vf ^a]$ is the Haar measure
for the hermitian matrix-valued field $H$ [1-4]. Here we parametrize $H$ in terms of
real fields $\vf^a (x)$ with
$H^{-1}dH = d\vf ^a r_{ak} (\vf) t_k$. $c_A$ is the quadratic Casimir of the
adjoint representation, $c_A \d^{ab} = f ^{amn} f^{bmn}$ and is equal to 
$N$ for an
$SU(N)$-gauge theory.
$\S (H)$ is the  Wess-Zumino-Witten
(WZW) action for the hermitian matrix field $H$ given by
$$  {\S} (H) = {1 \over {2 \pi}} \int \Tr (\partial H \bar{\partial} H^{-1}) +{i
\over {12 \pi}} \int \epsilon ^{\mu \nu \alpha} \Tr ( H^{-1} \partial _{\mu} H H^{-1}
\partial _{\nu}H H^{-1} \partial _{\alpha}H) \eqno(4)
$$
The inner product for the gauge-invariant physical states is given by
$$
\la 1 \vert 2\ra = \int  d\mu (H)  e^{2c_A \S (H)}~\Psi_1^* (H) \Psi_2 (H)
\eqno(5)
$$ 
This reduces matrix elements of the $(2+1)$-dimensional gauge theory
to Euclidean correlators of a hermitian WZW model. This is the essence of
the simplification of calculations in the gauge theory. Expressions for
the Hamiltonian, discussions of states and detailed analysis of
regulators are all given in references [1,2].

We now consider the reduction to gauge-invariant degrees of freedom
directly in the operator language, without choosing
a representation. The Gauss law operator $I^a(x)$ is given by
$$
I^a (\vx) = 2 (D \bE +\bD E )^a (\vx) \eqno(6)
$$
where $(D, \bD )$ are covariant derivatives, $Dh=\del h + [A,h ],~ \bD h= 
\bdel h +
[\bA, h]$. 
We shall consider $(\bE^a , I^a )$ as the independent variables writing
$$
E(\vx)= \int _y \bcalG(\vx, \vy) (\half I - D \bE )(\vy) \eqno(7)
$$
where $ (\bD_x \bcalG (\vx,\vy))^{ab} =\delta^{ab} \delta (\vx-\vy)$. The basic
commutation rules are
$$\eqalign{
[E^a (\vx), \bA^b (\vy) ] = [\bE^a (\vx) ,&A^b(\vy)] = -{\textstyle{i\over 2}}
\d^{ab}\d (\vx-\vy)\cr
[I^a(\vx), A^b(\vy)] &= -i D^{ab}_x \d (\vx-\vy)\cr}
\eqno(8)
$$
Eq.(7) is consistent with these commutation rules and hence is valid as an operator
identity. The kinetic energy operator can now be written as
$$
T= 2e^2 \int_x E^a (\vx) \bE^a (\vx)~= 2e^2 \int _{x,y} \left[
\bcalG^{ab}(\vx,\vy) \left( \half I -D\bE
\right)^b(\vy)\right]~\bE^a(\vx) \eqno(9)
$$
We now move the Gauss law operator to the right end of this expression. We find
$$\eqalign{
\half \int_y \bcalG^{ab}(\vx,\vy) I^b(\vy) \bE^a(\vx) &= \half \int_y 
\bcalG^{ab}(\vx,\vy) \bE^a(\vx)
I^b(\vy) -{\textstyle{i\over 2}}\int_{y} \bcalG^{ab}(\vx,\vy) f^{abc} \bE^c(\vy) 
\d (\vx-\vy)\cr
&= \half \int_y \bcalG^{ab}(\vx,\vy) 
\bE^a(\vx)
I^b(\vy) -\half \Tr \left[ T^c \bcalG (\vx,\vy) 
\right]_{\vy\rightarrow \vx}
\bE^c(\vx)\cr}
\eqno(10)
$$
where $T^c_{ab}= -if^{abc}$ is the adjoint representation of $t^c$.
The coincident point limit of the Green's function has to be evaluated
with a gauge-invariant regulator and, as we have noted before, it is equivalent
to an anomaly calculation in two Euclidean dimensions. (The Green's function
$\bcalG (\vx,\vy)$ can be considered as the propagator for a chiral fermion
in two Euclidean dimensions. The coincident point limit we need is the
fermionic current in a background field $A, ~\bA$. The covariant divergence of the
current is the standard gauge anomaly and hence we can obtain the current 
by integration of the anomaly.
Regularization issues have been discussed in detail in reference [2].) 
The result is
$$
- \half \Tr \left[ T^a \bcalG (\vx,\vy) \right]_{\vy\rightarrow \vx} = {ic_A\over 2\pi }
(A-M^{\dag -1} \del M^\dag )^a \eqno(11)
$$
(In terms of integrating the anomaly, this equation should read
$$
- \half \Tr \left[ T^a \bcalG (\vx,\vy) \right]_{\vy\rightarrow \vx}=
{ic_A\over 2\pi } \int_y \bcalG (\vx,\vy)^{ab} (\bdel A -\del \bA
+[\bA ,A])^b (\vy)
\eqno(12)
$$
A partial integration then leads to Eq.(11).)

Using Eqs.(10,11), the kinetic energy operator becomes
$$\eqalign{
T= 2im  \int_{x}& (A-M^{\dag -1} \del M^\dag )^a (\vx) \bE^a (\vx)~-~ 2e^2 
\int_{x,y} (\bcalG(\vx, \vy) D\bE (\vy))^a \bE^a (\vx)\cr
 ~&+~ e^2  \int_{x,y} 
\bcalG^{ab}(\vx,\vy) \bE^a(\vx)
I^b(\vy) \cr
}\eqno(13)
$$
where $m= {e^2c_A / 2\pi}$. On physical states which are annihilated by 
the Gauss law operator $I^a$, the last term gives zero. The first term
carries information about the mass gap.

Eq.(13) gives an expression for $T$ which is valid in both $E$- and
$A$-representations. In the $E$-representation, the quantity
$M^{\dag -1}\del M^\dag $ is a very nonlocal operator 
involving differentiations with respect to $E^a$. 
In the $A$-representation, we can simplify expression (13) further.
The parametrization (2) for the $A$'s can be written as
$$\eqalign{
A&= M^{\dag -1} (-\del H ~H^{-1} ) M^\dag ~+~ M^{\dag -1} \del M^\dag \cr
\bA &= M^{\dag -1}\bdel M^\dag \cr}\eqno(14)
$$
where $H=M^{\dag} M$. Thus $(A,\bA )$ is a complex $SL(N,{\bf C})$-gauge 
transform of $(-\del H~H^{-1},0)$.
Eventhough this involves a complex gauge transformation, it is possible to use 
this information to simplify $T$. In the $A$-representation, 
the wavefunction $\Psi (A,\bA )$ may be taken to be a function of $J^a= 
(c_A/\pi ) \del H~H^{-1}$ and $M^\dag$, as seen from (14). 
A change of $M^\dag$ is equivalent to a gauge transformation, 
but with complex gauge parameters. Thus we may write, for infinitesimal 
$\theta$,
$$
\Psi (M^\dag e^\theta ,J) \approx \Psi (M^\dag ,J) + \int \theta^a I^a 
~\Psi (M^\dag ,J)
\eqno(15)
$$
$I^a$ may be thought of as a functional differential operator on functions
of $J,M^\dag$. Eventhough $\theta$ is complex in general (and not purely 
imaginary as for a unitary transformation), the condition $I^a \Psi =0$ is 
sufficient to write
$$
\Psi (M^\dag e^\theta ,J) = \Psi (M^\dag ,J) \eqno(16)
$$
for physical states. By a sequence of such transformations, we may set
$M^\dag $ to $1$, i.e., $\Psi$ may be taken to be purely a function of $J$.
(Notice that, in two dimensions, all configurations $M^\dag$ can be
connected to the identity, i.e., are homotopic to the identity,
since $\Pi_2(SL(N,{\bf C}))=0$.) In this case, we may replace $A$ by
$-\del H~H^{-1}$, $\bA$ by zero and 
Eq.(13) for $T$ then becomes
$$\eqalign{
T&= m \left[ \int_u J^a(\vu) {\d \over \d J^a(\vu)} ~+~ \int \Omega_{ab} (\vu,\vv) 
{\d \over \d J^a(\vu) }{\d \over \d J^b(\vv) }\right]\cr
\Omega_{ab}(\vu,\vv)&= {c_A\over \pi^2} {\d_{ab} \over (u-v)^2} ~-~ 
i {f_{abc} J^c (\vv)\over {\pi (u-v)}}\cr}\eqno(17)
$$
Essentially the first term in $T$ provides a mass gap $\sim nm$ 
for a state composed out of
$n$ $J$'s. (Of course, this value may be modified by
extra contributions from the second term, see reference [2].)
 
In principle, one may also obtain the measure of integration for the inner product 
of the wavefunctions by requiring self-adjointness of the above expression.
This will coincide with the converse calculation in reference [2], where we
have checked that this expression is self-adjoint with the inner product
as given in Eq.(5).

Eq.(17) may be taken as the starting point for
analyzing the physical spectrum of the theory. Here we have
outlined a particular way to derive this expression. There are many
other ways to arrive at Eq.(17), some of
which are discussed in references [1,2].

In terms of the collective field variable $J$ the potential energy term is written as
$$
V = {1 \over {2 e^2}} \int B^2 (\vx) = { \pi \over {m c_A}} \int \bdel J_a (\vx)
\bdel J_a (\vx)
\eqno (18)
$$
Notice that, for momentum modes $k \ll e^2 \sim m $, the potential energy term gives
contributions of the order $k^2 / m$. For momenta of this order, 
$V$ can
be treated perturbatively. So in this regime, which can be thought of as a strong
coupling regime, one can in principle analyze the spectrum of the full theory by
studying the spectrum of the kinetic energy operator, Eq.(17), and including the
perturbative corrections from the potential energy term. This is the approach we are
going to follow in order to derive an expression for the vacuum wavefunction of the
theory. 

As far as the kinetic
energy operator is concerned, $\Psi_0 =1$ may be taken as the vacuum wavefunction. 
Trivial as it may seem, it is important that $\Psi_0 =1$ is 
normalizable with the inner product (5).
For low momentum modes, $k \ll m $ , the inclusion of the potential energy term leads 
to a modified vacuum wavefunction which can be written as
$$
\Psi = e^P \Psi _0
\eqno (19)
$$
where $P$ is a functional of the $J$'s which 
can be expanded in powers of $1/m$.
The various terms in this expansion can be determined from the Schr\"odinger 
equation for the vacuum wavefunction
$$
{\cal H} \Psi = (T +V) \Psi =0\eqno(20a)
$$
or equivalently
$$
\tilde{{\cal H}} \Psi_0 = e^{-P} (T+V) e^P \Psi_0 =0
\eqno(20b)
$$
Further, since $T$ contains at most two derivatives with respect
to $J$'s,
$$
\tilde{{\cal H}} = e^{-P} (T+V) e^P = T + V + [T, P] + \half [[T, P], P]
\eqno(21)
$$
Using Eqs.(17), (20) and (21), we can, in principle, 
calculate the full $1/m$-expansion of the vacuum
wavefunction. The first few terms are given as
$$ \eqalign{
P =  & - {\pi \over { m^2 c_A}} \Tr \int  : \bdel J \bdel J : \cr
& - \left({\pi \over { m^2 c_A}}\right)^2 \Tr \int   \bigl[: \bdel J ( {\cal D} \bd ) \bdel J 
    +  {1 \over 3} \bdel J  [J, \bdel ^2 J] : \bigr] \cr  
& - 2 \left({\pi \over {m^2 c_A}}\right)^3 \Tr \int \bigl[ : \bdel J  ( {\cal D} \bd )^2 \bd J 
+{2 \over 9} [ {\cal D} \bd J,~ \bd J] \bd ^2 J + {8 \over 9} [{\cal D} \bd ^2
J,~ J] \bd ^2 J \cr
&~~~~~~~~~~~~~~~~- {1 \over 6} [J, ~ \bd J] [\bd J,~ \bd ^2 J] - {2 \over 9} [J,
\bd J][J, \bd ^3 J]: \bigr]\cr 
& + {\cal O} ( {1 \over m^8})  \cr}
\eqno(22)
$$
where ${\cal D}h= {c_A \over \pi} \del h -[J,h]$. The normal ordering of various
terms in Eq.(22) is necessary for $P$ to satisfy Eq.(20). The second derivative
in Eq.(17) can give singularities when acting on composite operators. The normal
ordering subtracts out precisely these singularities.

There are several interesting points regarding the expansion in Eq.(22). 
The leading order
term for the vacuum wavefunction is
$$
\Psi \approx \exp \left[ -{\pi \over { m^2 c_A}} \Tr \int :\bdel J \bdel J :
\right] = 
\exp \left[ -{1\over 2me^2} \Tr \int B^2\right] \eqno(23)
$$
The calculation of expectation values involves averaging with the
factor $\Psi^* \Psi \approx e^{-S}$, where $S$, as seen from the above
equation, is the action of a Euclidean two-dimensional Yang-Mills
theory of coupling constant $g^2 = m e^2 = e^4 c_A /2 \pi $. 
Thus, retaining only the leading term in $\Psi$, 
the expectation value of the Wilson loop 
operator in the fundamental representation is given by
$$
\la W_F (C) \ra = \exp \left[ - {e^4 c_A c_F \over 4\pi } {\cal
A}_C \right]
\eqno(24)
$$
where ${\cal A}_C$ is the area of the loop $C$ and $c_F$ is the quadratic Casimir of the
fundamental representation [5]. The expectation value of the Wilson loop exhibits 
an area law behavior, as expected for a confining theory. 
{}From Eq.(24) we can easily identify the expression for
the string tension $\sigma $ as
$$
\sigma = {e^4 c_A c_F \over 4\pi } = e^4 \left( {N^2 -1\over 8\pi }\right)
\eqno(25)
$$

Recent Monte Carlo calculations of the string tension give the values
$\sqrt{\sigma}/e^2 =$ 0.335, 0.553, 0.758, 0.966  for the gauge 
groups $SU(2),~SU(3),
~SU(4)$ and $SU(5)$
respectively [6]. The corresponding values calculated from Eq.(25) are
0.345, 0.564, 0.772, 0.977. 
We see that there is excellent agreement (upto $\sim 3\%$) 
between Eq.(25)
and the Monte Carlo results. It is further interesting to notice that our
analytic expression for the string tension (25) has the appropriate
$N$-dependence as expected from large-$N$ calculations. 

Eq.(23) is roughly
in agreement with conjectures on the form of the vacuum wavefunction proposed
by Greensite and others [7, 8]. It was suggested there that,
for long wavelength configurations, the
vacuum wavefunction admits an expansion in terms of local 
gauge-invariant quantities of the form 
$$
\ln \Psi = \int {b_1 \over e^4} B^2 + {b_2 \over e^8} (D_i B)^2 + ...
\eqno(26)
$$
where $D_i$ is the covariant derivative, $D_i = \del _i - [A_i,~]$. Our 
analytical expansion (22) does not
quite agree with this conjecture. Our expansion is local in terms of the 
gauge-invariant
variables $J$, but not local in terms of $B$. 
One can easily work out the following relations between various derivatives of $J$ and $B$.
$$\eqalign{
\bdel ^n J & = - {c_A \over {2 \pi}} M^{\dag} ( \bar{D} ^{n-1} B) M^{\dag -1} \cr
( {\cal D} \bdel )^n \bdel J & = -{1 \over 2} \left({c_A \over \pi}\right)^{n+1} M^{\dag} 
(D \bar{D})^n B
M^{\dag -1} \cr}
\eqno(27)
$$
Using these relations one can easily check that all the expressions which involve only
derivatives of $J$ are local expressions in terms of $B$. The nonlocality appears in terms
which contain bare $J$'s. These terms involve the expression 
$M^{\dag -1} \del M^{\dag} - A $,
which is nonlocal in terms of $B$ since 
$ M^{\dag -1} \del M^{\dag} - A = - {1 \over 2} \bar{D} ^{-1} B $.
For example, the term of order $1/m^4$ in
Eq.(22) can be written in terms of $B$ as
$$
-{{c_A} \over {m^4 \pi}} \Tr \int \left[ {1 \over 4} B D \bar{D} B +
 {1 \over 24} [\bar{D} ^{-1} B, B
] \bar{D} B \right]
\eqno(28)
$$
The first term in the above expression is local in $B$ 
and is the same term that appears in
Eq.(26), but the second term is nonlocal in $B$. 
Similarly the term of order $1/m^6$ in Eq.(22) can be written in terms of $B$ as
$$\eqalign{
-{c_A \over {2m^6 \pi}} \Tr \int \bigl[
    & B (D\bar{D})^2 B - {1\over 9} [DB, B] \bar{D}B 
 - {4\over 9} [D\bar{D}B, \bar{D}^{-1}B] \bar{D}B
 - {1\over 24} [\bar{D}^{-1}B, B] [B, \bar{D}B]\cr
 & -{1\over 18} [\bar{D}^{-1}B, B] [\bar{D}^{-1}B, \bar{D}^2 B] \bigr]\cr}
\eqno(29)
$$

There have been several attempts to numerically estimate the coefficients
 $b_1,~b_2$ using
Monte Carlo simulations of the corresponding lattice gauge theory [8]. 
In these calculations a
local expansion as in Eq.(26) has been assumed. It is interesting to investigate 
whether one
could incorporate the nonlocal terms in lattice calculations.

The approximation of the vacuum wavefunction by the first few terms in Eq.(22) makes sense
only in the low momentum regime $k \ll m$. On the other hand if we were able to sum up the
whole series we could get information on the vacuum wavefunction away from the low momentum 
region. In fact, we can now show that Eq.(22) can be used to reconstruct the vacuum
wavefunction for short distances, $k \gg e^2$, which can be thought of as a weak coupling
regime. The terms in Eq.(22) can be naturally rearranged into terms with two
$J$'s, terms with three $J$'s, etc. This way we convert the $1/m$ expansion into
a series expansion in $J$'s. The series of terms with
only two $J$'s can be summed up to give
$$
P = -{1 \over {2 e^2}} \int_{x,y} B_a(\vx) \left[{ 1 \over  {\bigl( m + 
\sqrt{m^2 - \nabla ^2 } \bigr)} }\right] _{\vx,\vy} B_a(\vy)
\eqno(30)
$$
In the weak coupling or high momentum
regime $k \gg e^2$, the leading order term in Eq.(30) is
$$
P = - {1 \over {2 e^2}} \int_{x,y}~ B_a(\vx) \left[ 
(-\nabla^2)^{-\half}\right]_{\vx,\vy}  B_a(\vy)
\eqno(31)
$$
This is the vacuum wavefunction for an Abelian theory as expected.
Of course, in the low momentum regime $k \ll e^2$ the 
leading order term in 
Eq.(30) will
reproduce Eq.(23). (This wavefunction is similar to, but not quite the same, as the 
trial function suggested in reference [9].)

We now turn to the contribution of the 3$J$-terms. This can be, in
principle, derived by resumming all the 3$J$-terms in Eq.(22). An easier way
is to postulate a series expansion in $J$'s and solve the recursion relations on
the coefficients which follow from Eqs.(20,21).
This gives
$$
P  = - {2\over e^2}\left[
{{ \pi ^2} \over { {c_A}^2}} \int  \bd J_a \left[ { 1 \over {\bigl( m
+ \sqrt{m^2
-\nabla^2 } \bigr)}} \right] \bd J_a + f_{abc} \int f^{(3)}(\vx,\vy,\vec{z}) 
J_a(\vx) J_b(\vy) J_c(\vec{z})\right] 
\eqno(32)
$$
where $f^{(3)}$ is given, in momentum space, as
$$
f^{(3)}(\vk, \vp,\vq) =  (2\pi)^2 \delta({\vk}+{\vp}+{\vq})
   {1 \over 8}\left({\pi \over c_A}\right)^3
   {(E_k-m)(E_p-m) \over E_k + E_p + E_q} {\bar{k} - \bar{p} \over k p}
\eqno(33)   
$$
with $E_k = \sqrt{m^2 + \vk^2}$, etc. The momenta
in the denominator in
the above expression are the holomorphic components, $k= \half
(k_1+ik_2)$, etc. Some, but not all, of the non-Abelian terms involving
the structure constants are just what is needed to covariantize the
derivatives in the first term of Eq.(32), so that $\nabla^2$ is appropriately
changed to $4{\cal D}\bdel$, or equivalently $\nabla^2$ in Eq.(30)
is changed to the gauge-covariant Laplacian. 

The low momentum expansion of the terms in Eq.(32) reproduces
Eq.(22) to the appropriate order. At high momenta,
we see from
Eq.(33) that the $3J$-term is subdominant compared to
the leading $2J$-term in $P$. This is consistent with what 
is expected from perturbation theory. (In comparing with perturbation
theory, recall that $e A
\sim  ({\pi /c_A}) J$, where $A$ is the gauge potential. Therefore the $3J$-terms involve one power
of $e$ and an $f_{abc}$-factor.) This shows that the 3$J$-term contribution is
subdominant compared to the 2$J$-term in Eq.(32) for both the low and high
momentum regimes. Similar arguments hold, based on dimensional
analysis, for the higher $J$-terms. 
The analytic expansion in Eq.(32), as a series expansion in $J$'s  
is thus consistent with both the low and high momentum regimes.

So far we have discussed the structure of the vacuum wavefunction. One could, in principle, 
extend the above analysis for
the excitation spectrum of the Schr\"odinger equation (20). For example, in the absence of the 
potential term $V$,
the current $J$ is an eigenstate of the kinetic energy operator $T$, Eq.(17), with
eigenvalue $m$. One could now ask what the corresponding modified 
eigenstate and eigenvalue should be, once the
potential term is included. We would expect to find an expression of the form
$\tilde{J} e^P$, where $\tilde{J} = J + {\cal O}(J^2)$.
The higher $J$-terms can, in principle, be
calculated by solving the Schr\"odinger equation for $\tilde{\cal H}$,
where $\tilde{\cal H}$ is given by Eq.(21).  If we neglect the higher 
$J$-terms and keep only the 2$J$-term in the expression for $P$ we find
$$\eqalign{
\tilde{\cal H} =   \int & \left[ \sqrt{m^2 - \nabla ^2} ~J_a(\vx) \right] {\d \over 
{\d J_a (\vx)}} 
 + m \int 
\Omega_{ab} (\vx,\vy) 
{\d \over \d J^a(\vx) }{\d \over \d J^b(\vy) } \cr}
\eqno(34)
$$
As expected in a relativistic theory, the mass $m$ gets corrected to its 
relativistic
expression $\sqrt{m^2 + \vec{k} ^2}$. This is very similar to what happens with
solitons in a weak coupling expansion [10]. We are currently investigating how
the Hamiltonian gets modified by the inclusion of the higher $J$-terms in
Eq.(32). As in the case of the vacuum wavefunction we expect these terms to be
subdominant in both the low and high energy regimes.
\vskip .2in
\noindent{\bf Acknowledgements}
\vskip .1in

This work was supported in part by the DOE grant DE-FG02-91ER40651-Task B and
NSF grant PHY-9605216. C.K.'s work was supported by the Korea Science and
Engineering Foundation through the SRC program.
\vskip .2in
\noindent{\bf References}
\vskip .1in
\item{1.}
D. Karabali and V.P. Nair, {\it Nucl.Phys.} {\bf B464} (1996) 135; {\it Phys.Lett.}
{\bf B379} (1996) 141; {\it Int. J. Mod. Phys} {\bf A12} (1997) 1161.
\item{2.}
D. Karabali, C. Kim and V.P. Nair, hep-th/9705087, to appear in {\it Nucl.
Phys.} {\bf B}.
\item{3.}
A.M. Polyakov and P.B. Wiegmann, {\it Phys.Lett.} {\bf B141} (1984) 223;
D. Karabali, Q-H. Park, H.J. Schnitzer and Z. Yang, {\it Phys. Lett.} {\bf B216}
(1989) 307; D. Karabali and H.J. Schnitzer, {\it Nucl. Phys.} {\bf B329} (1990)
649.
\item{4.} 
K. Gawedzki and A. Kupiainen, {\it Phys.Lett.} {\bf B215} (1988) 119;
{\it Nucl.Phys.} {\bf B320} (1989) 649;
M. Bos and V.P. Nair, {\it Int.J.Mod.Phys.} {\bf A5} (1990) 959.
\item{5.} 
A. Migdal, {\it Zh.Eksp.Teor.Fiz.} {\bf 69} (1975) 810  ({\it Sov.Phys.JETP}
{\bf 42} (1975) 413); B. Rusakov, {\it Mod.Phys.Lett.} {\bf A5} (1990) 693; E. Witten,
{\it Commun.Math.Phys.} {\bf 141} (1991) 153; D. Fine, {\it Commun.Math.Phys.} {\bf 134}
(1990) 273; M. Blau and G. Thompson, {\it Int.J.Mod.Phys.} {\bf A7} (1992) 3781; D.
Gross, {\it Nucl.Phys.} {\bf B400} (1993) 161; D. Gross and W. Taylor, {\it Nucl.
Phys.} {\bf B400} (1993) 181; J. Minahan, {\it Phys.Rev.} {\bf D47} (1993) 3430.
\item{6.} 
M. Teper, {\it Phys. Lett} {\bf B311} (1993) 223; hep-lat/9804008 and references therein.
\item{7.} 
J. Greensite, {\it Nucl. Phys.} {\bf B158} (1979) 469; 
M.B. Halpern, {\it Phys. Rev.} {\bf D19} (1979) 517; P. Mansfield, {\it Nucl.
Phys.} {\bf B418} (1994) 113.
\item{8.}
J. Greensite, {\it Phys.Lett.} {\bf B191} (1987) 431;
H. Arisue, {\it Phys.Lett.} {\bf B280} (1992) 85; Q.Z. Chen, X.Q. Luo and S.H.
Guo, {\it Phys. Lett.} {\bf B341} (1995) 349.
\item{9.} 
S. Samuel, {\it Phys. Rev.} {\bf D55} (1997) 4189. 
\item{10.}
see, for example, B. Sakita, {\it Quantum theory of many variable systems and fields}
(World Scientific, 1985).

\end